# The gamma-ray burst GB 920622

J. Greiner[1], M. Sommer[1], N. Bade[2], G.J. Fishman[3], L.O. Hanlon[4], K. Hurley[5], R.M. Kippen[6], C. Kouveliotou[3], R. Preece[3], J. Ryan[6], V. Schönfelder[1], O.R. Williams[4], C. Winkler[4] M. Boer[7], M. Niel[7]

[1] Max-Planck-Institut für Extraterrestrische Physik, 85740 Garching, Germany
[2] Hamburg Observatory, 21029 Hamburg, Germany
[3] Marshall Space Flight Center, Huntsville, AL 35812, U.S.A.
[4] Astrophysics Division, ESA-ESTEC, 2200 AG Noordwijk, The Netherlands
[5] Space Science Laboratory, University of California, Berkeley CA 94720, U.S.A.
[6] Space Science Center, University of New Hampshire, Durham, NH 03824, U.S.A.
[7] Centre d'Etude Spatiale des Rayonnements, 31029 Toulouse, France

**Abstract.** We have analyzed the Ulysses, BATSE, and COMPTEL spectral data from the $\gamma$-ray burst of June 22, 1992 (GB 920622). COMPTEL data reveal a hard to soft evolution within the first pulse of the burst, while the mean hardness ratios of the three pulses are the same. Unlike the single instrument spectra, the composite spectrum of GB 920622 averaged over the total burst duration ranging from 20 keV up to 10 MeV cannot be fit by a single power law. Instead, the spectrum shows continuous curvature across the full energy range.

COMPTEL imaging and BATSE/Ulysses triangulation constrain the source location of GB 920622 to a ring sector 1.1 arcmin wide and 2 degrees long. This area has been searched for quiescent X-ray sources using ROSAT survey data collected about two years before the burst. After the optical identification of the X-ray sources in and near the GRB location we conclude that no quiescent X-ray counterpart candidate for GB 920622 has been found.

## 1. Introduction

Gamma-ray bursts are recorded by all four instruments onboard the Compton Gamma-Ray Observatory (CGRO). However, apart from the Burst and Transient Source Experiment (BATSE), the most interesting data are collected when the burst happens to be in the field of view (FOV) of the other instruments (1 sr for the Compton Telescope (COMPTEL), 0.5 sr for the Energetic Gamma-Ray Experiment Telescope (EGRET), and $11^\circ \times 4^\circ$ for the Oriented Scintillation Spectrometer Experiment (OSSE)). Although only a few bursts per year among the ones which occur in the COMPTEL/EGRET FOV are intense and hard enough at these high energies to result in significant detections, the data from these events are extremely valuable in two respects: 1) Since the energy ranges overlap, these bursts allow a cross-correlation of the different instruments with their completely different measurement principles and deconvolution methods (Schaefer et al. 1994). 2) The spectra of these bursts can be determined over an unprecedented range in energy, from about 20 keV up to a few GeV (if the burst contains such energetic photons). This may allow us to draw unique conclusions about the emission mechanism in the bursts. In the two bursts 1B 910503 (Schaefer et al. 1994) and 1B 910601 (Share et al. 1994) investigated previously, the broad-band spectral shape is not compatible with single component models even though the single instrument spectra could be well fitted with such simple models.

In this investigation we present the CGRO data for the $\gamma$-ray burst of June 22, 1992. Since this burst was outside the OSSE and EGRET FOV, we are dealing with BATSE, COMPTEL and the EGRET anticoincidence dome data (section 2). In addition, we have included the low-energy data of the burst detector onboard the Ulysses spacecraft. For the first time, we have fitted simultaneously Ulysses, BATSE and COMPTEL count rate spectral data with the corresponding detector response matrices (section 3.2). Furthermore, using BATSE/COMPTEL/Ulysses data we derive the position of the burst (section 3.3) with an accuracy which has allowed a search for quiescent X-ray sources in the ROSAT All-Sky-Survey data (section 4).

## 2. Instruments

The BATSE instrument consists of eight modules (Fishman et al. 1989) each, placed at the corners of CGRO.

a Spectroscopy Detector (SD). Photons are detected in 128 LAD energy channels covering the energy range between ∼25 keV and ∼2 MeV. In addition there are four broad discriminator channels: 25–50 keV, 50–100 keV, 100–300 keV and >300 keV. A burst trigger is generated by the BATSE on-board software if two or more detectors measure a >5.5$\sigma$ increase in count rate in any of three time intervals (64 ms, 256 ms or 1024 ms) over the average background count rate of the preceding 17 sec. The data presented below are derived only from the LADs.

The OSSE detector utilizes four actively shielded and passively collimated NaI scintillation detectors with a $3°.8 \times 11°.4$ FWHM FOV. The four annular NaI shields surrounding the OSSE detectors continuously accumulate data with a time sampling interval of typically 16 ms, which are dumped only in response to a BATSE burst trigger. There is no pulse height analysis of the detected photons and the low-energy threshold is about 150 keV.

The COMPTEL telescope onboard CGRO operates in the 0.75 to 30 MeV range with a field of view of about 1 steradian (Schönfelder et al. 1993). Two different modes of operation are employed by COMPTEL for the detection of GRBs (Winkler et al. 1986). In the "telescope" mode (which is the usual imaging mode of COMPTEL), each incoming $\gamma$-ray photon is first Compton-scattered in the upper layer of detectors (D1) and then absorbed in the lower detector layer (D2). This allows images to be produced, and spectra and light curves of GRBs occurring in the field of view of COMPTEL to be measured. In the "burst" mode (BSA data), which is triggered upon receipt of a signal from BATSE, two of the lower layer NaI detectors (called D2-7 and D2-14) accumulate 6 spectra for an integration time of 1 sec each ("burst" spectra), followed by 133 spectra of 6 sec integration time ("tail" spectra). The two burst modules operate in two overlapping energy ranges: 0.3–1.3 MeV (D2-14) and 0.6–10 MeV (D2-7). The spectral resolution is 9.6% at 0.5 MeV and 7% at 1.5 MeV. Different data selections for spatial, spectral and timing analysis have been applied to give results less affected by instrument characteristics.

The EGRET total absorption shower counter (TASC) is a NaI scintillation detector measuring $77 \times 77 \times 20$ cm$^3$ with an axial thickness of about 8 radiation lengths. Its normal objective is to determine the energy of gamma-ray photon showers which converted in the spark chamber telescope above the TASC. In a secondary mode, the spectrum of the energy deposit by any radiation (i.e. no charged particle discrimination and no veto counter dead time losses) in the TASC is recorded in the 0.6 to 170 MeV range with two types of accumulation times (Kanbach et al. 1988): a regularly repeated mode provides spectra every 32 sec and a special burst mode gives spectra with adjustable time intervals (typically 1, 2, 4 and 8 sec) after a trigger from the BATSE instrument.

25–150 keV energy band (Hurley et al. 1992). Two hemispherical CsI detectors of 20 cm$^2$ effective projected area provide a $4\pi$ steradian field of view. The duty cycle of the burst detector has been >95% since November 1990.

## 3. The gamma-ray burst of June 22, 1992

The BATSE detectors on CGRO triggered on a strong $\gamma$-ray burst on June 22, 1992 (GB 920622) at T$_{trig}$(UT) = $07^h05^m04^s.6$ = 25504.6 seconds (BATSE trigger number 1663). GB 920622 occurred within the COMPTEL field of view (though at a relatively large off-axis angle of 45°.5), thus enabling imaging of this burst. Both burst modules (D2-7 and D2-14) were operational at the time of the burst and registered "burst" and "tail" spectra. The burst is also seen in the shields of the OSSE and EGRET detectors. GB 920622 was strong enough below 150 keV to trigger the burst detector on Ulysses, which recorded 1 sec duration spectra as well as the light curve in the 25–150 keV range with 1/32 sec temporal resolution.

### 3.1. Time History

In the BATSE LAD energy range (25 keV–2 MeV) the burst had a T$_{90}$ duration (see Fishman et al. 1994 for definition of T$_{90}$) of 50 sec with a 1 sec rise time, complex structure and an exponential decay to background (Fig. 1). There was weak precursor emission at T$_{trig}$–50 sec (Fig. 2). The peak count rate seen by BATSE occurred at T$_{trig}$+16 and significant emission above 300 keV was detected as can be seen in the COMPTEL data (Fig. 1). There are three consecutive pulses of emission each lasting about 5 sec. This is also clearly seen in the OSSE data (publicly available via XMOSAIC, Matz et al. 1994). Fig. 1 shows the burst time history at energies >150 keV as recorded by the summed shield elements of OSSE.

The binned COMPTEL "telescope" light curve in the energy range ∼0.7–10 MeV (Fig. 1) shows emission starting just after the BATSE trigger time and lasting for about 20 sec. The data selection was standard (see Schönfelder et al. 1993) except for the relaxation to D1E = D2E = ETOT = 0–90000 and phibar = 0–180°. Only events from within 10 degrees of the source were used for the light curve. This gives 26, 52, and 63 telescope events in the first, second and third pulse or a total of 141 events. The three pulses are also seen clearly. The most intense pulse has a rise time of ∼200 msec. Separating the data into three energy bands suggests that the third pulse appears to be relatively stronger in the higher energy bands than the other two pulses. However, detailed analysis is impossible due to low count statistics.

The accumulation times of the two COMPTEL burst modules D2-7 and D2-14 were set at 1 sec in burst mode (6 intervals after burst trigger), 6 sec in tail mode and 100 sec in background mode. The last complete 100 sec in-used for background subtraction. Fig. 1 shows the burst time history after background subtraction as seen by the two burst modules. Concentrating only on the first pulse measured at 1 sec resolution, the comparison of the relative intensities in the hard and soft energy bands suggests that there is spectral evolution within this first pulse of the burst (see section 3.2.1).

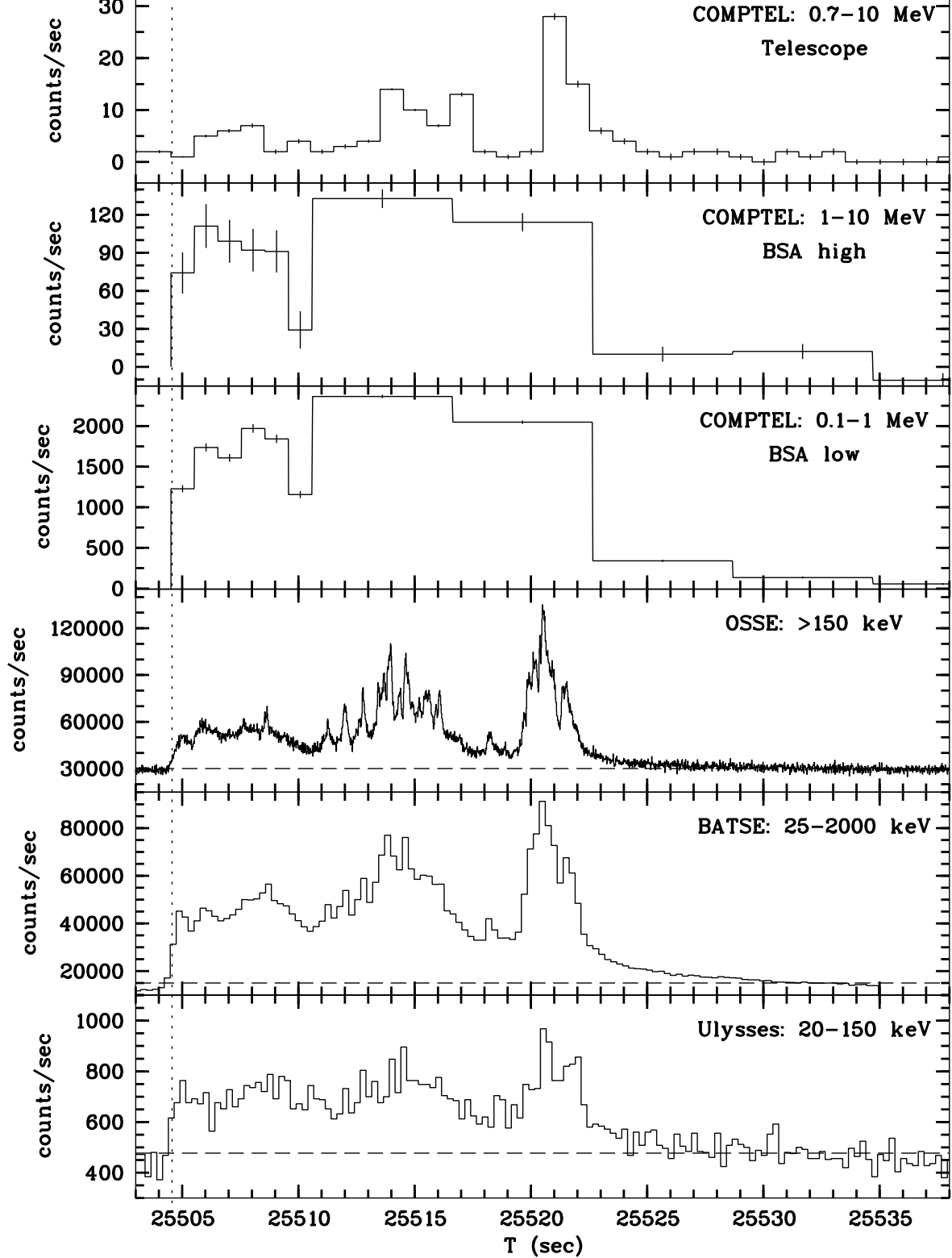

**Fig. 1.** Light curve of GB 920622 as seen by different instruments in different energy bands. Only the COMPTEL light curves are background subtracted. For the other detectors the dashed line gives the background rate. In most of the panels the error bars are smaller than the line thickness. The BATSE trigger time at 25504.568 sec is marked by the vertical dotted line. Note the different relative shapes at 4–6 sec after the trigger time (end of first pulse).

The Ulysses detector recorded the light curve with a time resolution of 1/32 sec, including 8 sec of pre-trigger data. The precursor emission was too weak to be detectable by either COMPTEL or Ulysses.

### 3.2. Spectrum

#### 3.2.1. Fitting to single instrument data

The LAD 4 BATSE data have been used with channels 0–22 (24–95 keV) excluded from the spectral analysis due to low-energy inaccuracies of the calibration. Several models were applied to the total burst spectrum (25.9 sec accumulation time after BATSE trigger) with a broken power law giving a not acceptable fit ($\chi^2/\nu = 188/89$). A better $\chi^2/\nu$ (115/89) is obtained using the "GRB model" introduced by Band et al. (1993) (see below). Details are given in Table 1.

Recently, Band et al. (1993) introduced the following mathematical model for fitting BATSE SD GRB spectra:

$$N_E(E) = A \left[\frac{E}{(100keV)}\right]^{\alpha} \exp\left[-\frac{E}{E_o}\right], \quad (\alpha - \beta)E_o \geq E$$

$$= A \left[\frac{(\alpha - \beta)E_o}{(100keV)}\right]^{\alpha - \beta} \exp(\beta - \alpha) \left[\frac{E}{(100keV)}\right]^{\beta}$$

$$(\alpha - \beta)E_o \leq E$$

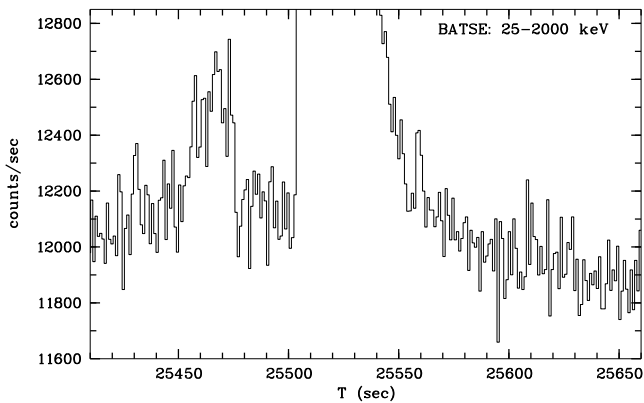

**Fig. 2.** BATSE light curve of GB 920622 showing the weak precursor about 50 sec before the trigger time. There is no emission above 300 keV in the precursor.

ing, it allows different standard models to be reproduced by using specific values of $\alpha$, $\beta$ or $E_o$, e.g. a simple power law ($E_o = \infty$) or the common burst continuum description ($\alpha \sim -1$, $\beta \sim -2$). In general, the low-energy part of the spectrum (20–100 keV) determines the value of $\alpha$, whereas the high-energy part determines $\beta$ (the $E^{\beta}$ power law is referred to as the high-energy tail in Band et al. 1993).

The data selection for the COMPTEL telescope data was the same as in the temporal analysis. The telescope livetime over the full $\sim$24 sec burst duration is estimated to be about 92%. Due to the limited temporal resolution of onboard livetime data we cannot make reliable livetime estimates for short, sub-burst intervals.

We have used the Cash test (Cash 1979) for fitting the COMPTEL telescope spectra of both, the full burst and of burst sub-intervals. Several different spectral models were tested, including power law, optically thin thermal bremsstrahlung (OTTB), Synchrotron, Compton, and broken power law. For the full burst interval, the simple power law model and models with more free parameters (Compton, broken power law) gave acceptable fits while simple models with curvature (OTTB, Synchrotron) gave worse results. Details of the deconvolution with the best fit power law model are given in Table 1.

The standard energy ranges for spectral fitting of the burst module data (Hanlon et al. 1994) are 0.3–1.6 MeV (D2-14) and 0.6–10 MeV (D2-7). All energy channels are rebinned to have the same signal/noise ratio of 3$\sigma$. Although GB 920622 was $\approx$45° off-axis it was possible to use an on-axis response matrix for the spectral fitting after the application of a simple correction factor of 1.4 for low range data and 1.2 for high range data (derived from pre-launch calibration data). A power law model was fit to the time integrated burst spectrum as well as to individual spectra. In all cases, there was no need for two component models (like broken power law) or models with more parameters (like the Band model) due to the goodness of the power law fit. The results are shown in Table 1. The errors are 90% (=1.6$\sigma$) errors on the two-parameter model (powl) and thus are the most conservative error estimates. The power law spectral index is plotted as a function of time in Fig. 3. In addition, hardness ratios have been calculated for each time interval individually and plotted in the same figure. Consistently, both show evidence for spectral softening within the first pulse of the burst. The average hardness ratio of the six 1 sec "burst" spectra is similar to the hardness ratio of the first and second "tail" spectra as well as to the average of the full burst interval. This means that there is certainly no spectral evolution at MeV energies over the full burst, i.e. from pulse to pulse. Summarizing, the spectrum of GB 920622 in the COMPTEL range may be represented as a single power law, with no evidence for spectral breaks or turnovers.

The Ulysses instrument collects 16 channel spectra with a time resolution between 1 and 16 seconds. The

**Table 1.** Spectral fitting results for single instrument data

| Instrument / Data type | Model | Time Interval | Fit parameters | | $\chi^2/\nu$ |
|---|---|---|---|---|---|
| | | | Norm* | Others | |
| COMPTEL Telescope | powl | Full burst | $1.01^{+0.17}_{-0.16}$ | $\alpha=-2.69^{+0.28}_{-0.30}$ | 0.26** |
| COMPTEL Telescope | powl | Pulse 1 | $0.75^{+0.32}_{-0.25}$ | $\alpha=-2.75^{+0.63}_{-0.75}$ | 0.47** |
| COMPTEL Telescope | powl | Pulse 2 | $1.17^{+0.33}_{-0.28}$ | $\alpha=-2.92^{+0.47}_{-0.52}$ | 0.85** |
| COMPTEL Telescope | powl | Pulse 3 | $1.08^{+0.29}_{-0.25}$ | $\alpha=-2.58^{+0.39}_{-0.30}$ | 0.28** |
| COMPTEL BSA low | powl | Full burst | $0.64^{+0.04}_{-0.04}$ | $\alpha=-2.46^{+0.07}_{-0.08}$ | 100/72 |
| COMPTEL BSA high | powl | Full burst | $0.70^{+0.04}_{-0.04}$ | $\alpha=-2.43^{+0.1}_{-0.13}$ | 26/25 |
| BATSE LAD | broken powl | 25.9 sec | $(5.0\pm0.04)\times10^{-2}$ | $E_{break}=366\pm10$ keV $\alpha_1=-1.37\pm0.01$ $\alpha_2=-2.27\pm0.02$ | 188/89 |
| BATSE LAD | Band | 25.9 sec | $(5.9\pm0.06)\times10^{-2}$ | $E_p=476\pm15$ keV $\alpha=-0.91\pm0.04$ $\beta=-2.38\pm0.05$ | 115/89 |
| Ulysses | powl | Full burst | $2.7\pm0.3$ | $\alpha=-0.8\pm0.3$ | 19/12 |

* Note that the normalizations are defined differently: for the COMPTEL data the normalization is in ph cm$^{-2}$ s$^{-1}$ MeV$^{-1}$ at 1 MeV; for the models applied to the BATSE data it is in ph cm$^{-2}$ s$^{-1}$ keV$^{-1}$ at 100 keV; and for the power law model applied to the Ulysses data it is in ph cm$^{-2}$ s$^{-1}$ keV$^{-1}$ at 1 keV.

** Cash test used, there is no $\chi^2$. The goodness-of-fit was determined by bootstrap simulations. The parameter of interest is P = probability of exceeding the observed best-fit Cash statistic purely by chance fluctuations of the model (this is completely analogous to $\chi^2/\nu$). In this case, one can reject models with (1-P) confidence; thus acceptable model fits have P $\approx$0.5 which directly corresponds to $\chi^2/\nu=1$.

timing of the spectra is such that the first spectrum is accumulated over a long time prior to the trigger, while the second may be accumulated over a very short time interval; these two spectra have been omitted in the analysis. A dead time correction of 22.6% has been applied to the count rate data. The spectrum averaged over 14 sec (starting 3.8 sec after the trigger) is consistent with a power law model (Table 1), but has a considerably flatter slope than the power law model fitted at MeV energies (COMPTEL and BATSE). The first two channels (< 17.5 keV) have been ignored in the fit.

### 3.2.2. Combined fitting of Ulysses/BATSE/COMPTEL count rate spectra

Inspecting the combined Ulysses/BATSE/COMPTEL spectrum of GB 920622 (Fig. 4) there are two things to note. 1) Given the completely different measurement principles and deconvolution algorithms of the three instruments involved, we find that the normalizations as well as the global spectral shape are in remarkable agreement in the overlapping parts. This is most clearly seen in the bottom panel of Fig. 4 where the differential spectral points are multiplied by $E^2$. 2) There is clear evidence that the composite broad-band spectrum cannot be fit by a single power law model. Instead, the continuum shows continuous curvature over the total range; thus appropriate models with curvature have to be used for fitting.

The count rate spectra and the detector response matrices of Ulysses, BATSE and COMPTEL (burst module data only) have been converted into XSPEC format and all three instrument data have been fitted simultaneously. Since the spectra were accumulated over different integration times (14 sec – with 3.8 sec offset – for Ulysses, 25.9 s for BATSE, 24 s for COMPTEL and 32 s for EGRET) defined (section 2) by the instrument set-up, correction factors were applied to the count rate spectra.

Again, several models have been tested and a summary is given in Tab. 2 and plotted in Fig. 4. Band's GRB model gives the fit with the lowest reduced $\chi^2$, and the following parameters are derived (errors are 90% confidence for 1 parameter of interest): $\alpha=-0.86\pm0.15$, $\beta=-2.51\pm0.07$, $E_o=457\pm30$ keV with a normalization amplitude of $6.0\times10^{-2}$ ph cm$^{-2}$ s$^{-1}$ keV$^{-1}$ at 100 keV. We have included a 10% systematical error in the absolute normalization of the COMPTEL burst module data. The

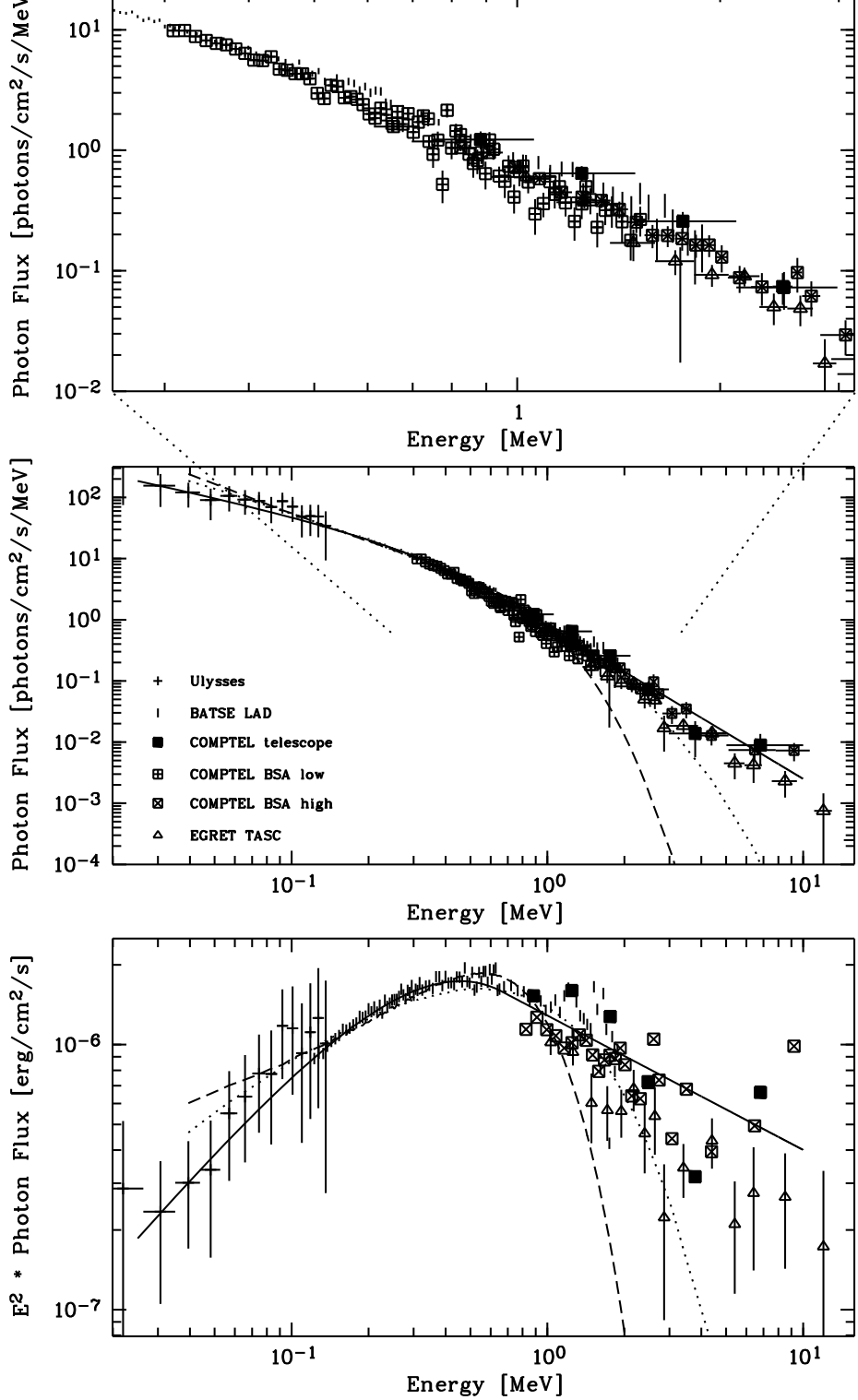

**Fig. 4.** Mean composite Ulysses, BATSE, COMPTEL and EGRET spectrum of GB 920622 over the full burst duration as seen in each instrument (see text). The errors are $1\sigma$ statistical errors for the Ulysses, BATSE and COMPTEL telescope data with 10% systematic error included for COMPTEL burst module data. The EGRET TASC data are from Schneid et al. (1995) and were not included in the fit procedure. Lines show the best fit models (see Table 2) to the Ulysses, BATSE, and COMPTEL data: solid line – Band model, dotted line – bremsstrahlung model, dashed line – Comptonization model after Sunyaev/Titarchuk. In the lower panel the COMPTEL low range data have been omitted for clarity.

**Table 2.** Results of the spectral fitting of the combined Ulysses/BATSE/COMPTEL data

| Model | Fit parameters | | $\chi^2/\nu$ |
|---|---|---|---|
| | Norm ph cm$^{-2}$ s$^{-1}$ keV$^{-1}$ | Others | |
| Band | $(6.0\pm0.3)\times10^{-2}$ | $E_p = 457\pm30$ keV $\alpha = -0.86\pm0.15$ $\beta = -2.51\pm0.07$ | 248/210 |
| Bremsstrahlung | $2.7\pm0.3$ | kT $= 849\pm25$ keV | 412/210 |
| Comptonization | $105\pm13$ | kT $= 208\pm10$ keV $\tau = 1.2\pm0.1$ | 620/210 |

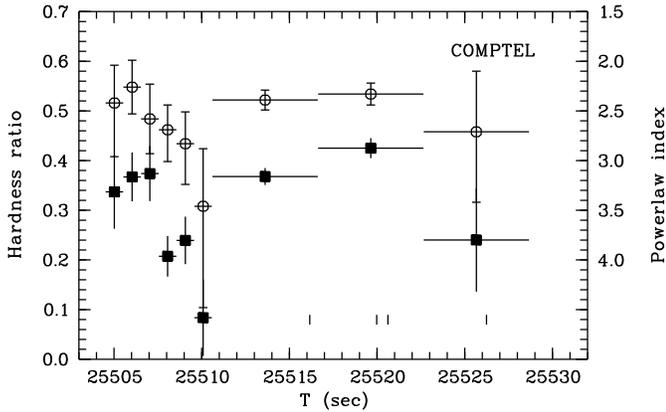

**Fig. 3.** Spectral evolution of the burst as seen by the COMPTEL burst modules. Shown are the hardness ratio (0.7–1.7 MeV) / (0.3–0.7 MeV) (filled squares, left axis) and the photon indices of single power law spectral fits to the BSA data (open circles, right axis) vs. time of day 1992 June 22. The asymmetric error of the fitted power law index is marked by the small cross bar at the end of the error bar. The error of the hardness ratio (1$\sigma$ plotted) is symmetric. Within the first pulse of the burst, i.e. during the first 6 seconds, significant softening is observed. The four small vertical bars indicate the arrival times of the photons with energies greater than 4 MeV detected in the COMPTEL telescope mode.

**Table 3.** Fluence of GB 920622

| Energy range | Fluence* |
|---|---|
| 20 keV – 10 MeV (combined fit) | $1.4\times10^{-4}$ erg cm$^{-2}$ |
| 30 keV – 2 MeV (KONUS range) | $1.1\times10^{-4}$ erg cm$^{-2}$ |
| 0.3 – 10 MeV (SMM range) | $0.9\times10^{-4}$ erg cm$^{-2}$ |

* A mean integration time of 25 sec has been used.

### 3.3. Localization

measured fluence of the burst is $1.4\times10^{-4}$ erg/cm$^2$ above 20 keV for a mean duration of 25 sec (see Tab. 3 for values using different energy bands). A high temperature ($\approx$100 keV or higher) blackbody model does not fit at all.

EGRET TASC data of GB 920622 have been fitted separately with a $-3$ photon index power law model (Schneid et al. 1995) and are added in Fig. 4. These EGRET data follow the extrapolation of the best fit exponential cut-off spectrum of Ulysses, BATSE and COMPTEL data below 1 MeV.

The accuracy of the most likely position derived from the BATSE data (as shown in Fig. 5 and Table 4) is primarily determined by systematic uncertainties since the 1$\sigma$ statistical error is only 0°.2 due to the strength of the burst. An independent position determination of the precursor emission gives a location consistent with that of the main burst emission.

For the COMPTEL telescope mode data reduction the module D2-2 was excluded from the imaging analysis due to a non-operating central photomultiplier tube. We used a one degree (chi, psi) binning and a two degree phibar binning in the range $0 \leq$ phibar $\leq 50°$ (see Schönfelder et al. 1993 for details on these dataspace variables). Data for the first 24 sec after the BATSE trigger were chosen, and the following data selections applied: TOF: 110–130, PSD: 0–110, D1E: 70–2000 keV, D2E: 650–30000 keV, ETOT: 720–30000 keV. This finally resulted in 162 events used for imaging.

Since this burst was far off-axis in the COMPTEL FOV, a simulated point spread function was generated specifically for this burst position, with a spectral shape derived from the burst module data analysis (E$^{-2.5}$ power law). Maximum entropy and maximum likelihood analysis algorithms both give a centroid position of $\alpha(2000.0) = 10^h\ 51^m\ 12^s$ and $\delta(2000.0) = 48°.0$. The 3$\sigma$ positional uncertainty is 1°.5 by 6° with the asymmetry due to the large off-axis angle. This position is fully consistent with the BATSE location at $\alpha(2000.0) = 10^h\ 26^m$, $\delta(2000.0) = 45°.0$ (given the $4°$ uncertainty).

**Table 4.** Position of GB 920622 (all equinox 2000.0)

| | BATSE |
|---|---|
| centroid position | $\alpha = 10^h\ 30^m\ (157°.5)$ |
| | $\delta = 44°.4$ |
| error radius | $4°.2$ |

| | COMPTEL |
|---|---|
| centroid position | $\alpha = 10^h\ 51^m\ 12^s\ (162°.8)$ |
| | $\delta = 48°.0$ |

| | Ulysses/BATSE triangulation arc |
|---|---|
| Center | $\alpha = 10^h\ 00^m\ 40^s\ (150°.167)$ |
| | $\delta = 6°.667$ |
| radius | $41°.333$ |
| width | $1'.1$ |

| | Ulysses/BATSE/COMPTEL |
|---|---|
| best position along arc | $\alpha = 10^h\ 46^m\ 46^s\ (161°.69)$ |
| | $\delta = 46°.80$ |
| $2\sigma$ error (at 189.44) | $\triangle\alpha = ^{+0°.46}_{-1°.39}, \triangle\delta = ^{-0°.10}_{+0°.28}$ |
| maximum arc likelihood | 193.4 |

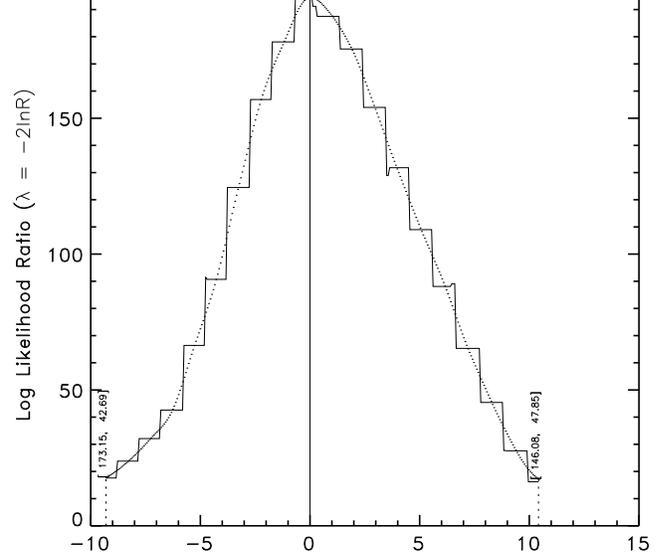

**Fig. 6.** COMPTEL maximum likelihood ratio evaluated along the triangulation arc as a function of angular distance (in degrees) from the position of the maximum likelihood ($\alpha = 10^h\ 46^m\ 46^s$ (161°.69), $\delta = 46°.80$). The solid line marks the measured likelihood values, and the dotted line is a bicubic interpolation of the likelihood map.

The Ulysses GRB detector recorded the burst 2173.282 sec earlier than the instruments on CGRO. At that time, the separation between the two spacecraft was 2894.381 light seconds. The resulting timing arc has a width of about $1'.1$ due to an intersatellite timing uncertainty of $\pm 0.3$ sec. The arc passes through the COMPTEL map at the $1.4\sigma$ level at its closest approach. Evaluating the COMPTEL likelihood ratio map at points along the triangulation arc results in a $2\sigma$ error box measuring less than $2°$ around the maximum point at $\alpha(2000.0) = 10^h 46^m 45^s.6$ (161°.69) and $\delta(2000.0) = 46°.8$ (Fig. 6 and Table 4).

## 4. Counterpart search

### 4.1. Soft X-ray sources

The X-ray satellite ROSAT (Trümper 1983) performed an All-Sky-Survey between August 1990 and January 1991. This was the first time that the whole sky was scanned in the soft X-ray range (0.1–2.4 keV) using a position sensitive proportional counter (PSPC, Pfeffermann et al. 1986).

The area of GB 920622 was scanned by ROSAT from Oct. 31, 1990 until Nov. 11, 1990 for a total observing time per location of about 540 sec, resulting in the detection of about 1.6 sources with likelihood greater than $5\sigma$ per square degree. Thus, within the $2\sigma$ error box size of 132 arcmin$^2$ one would expect 0.05 background X-ray source.

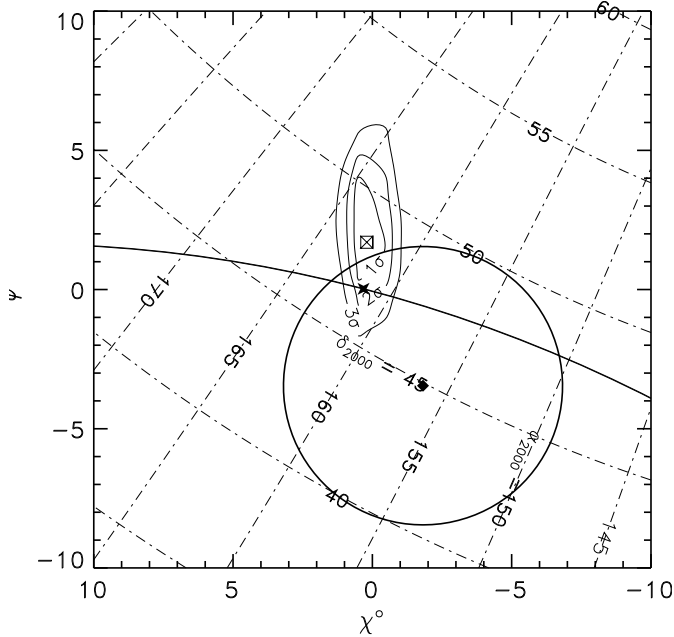

**Fig. 5.** COMPTEL maximum likelihood image of GB 920622 with equatorial coordinates (equinox 2000.0) superimposed (dash-dot line). The likelihood contours are at the $1\sigma$, $2\sigma$ and $3\sigma$ confidence level and the most probable likelihood position is marked with a squared cross. The curved line gives the BATSE/Ulysses triangulation arc. The small star on top of the triangulation arc is the best COMPTEL position along the arc (corresponding to the maximum of Fig. 6). The circle gives the BATSE position with a $5°$ error radius around the centroid position marked by a filled lozenge.

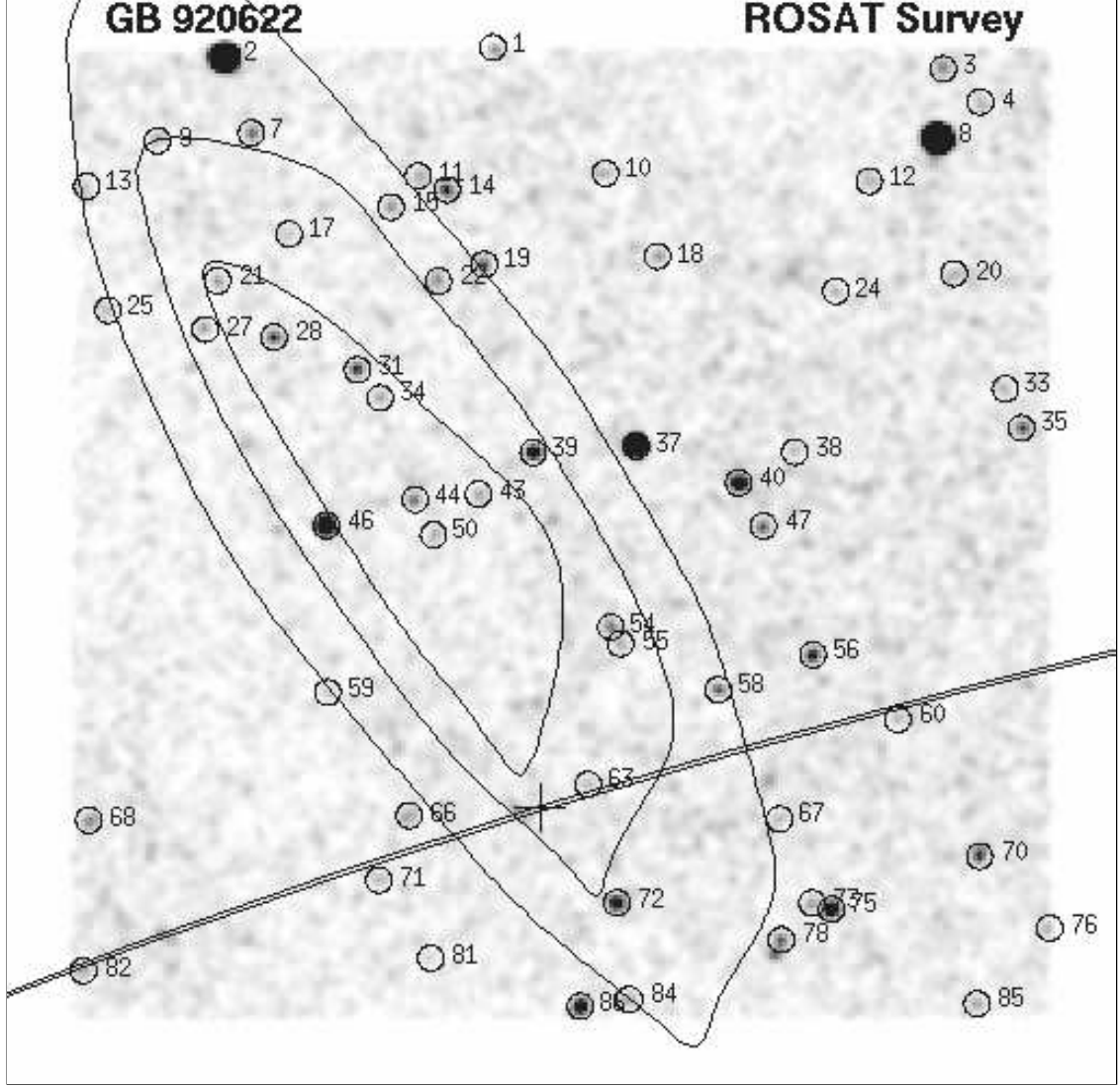

**Fig. 7.** Smoothed ROSAT survey image ($6° \times 6°$) with the COMPTEL error box of GB 920622 and the $1\rlap{.}'1$ wide BATSE/Ulysses triangulation arc superimposed. The cross denotes the best COMPTEL position along the triangulation arc. Small circles mark soft X-ray sources above $5\sigma$ with the color density within these circles being proportional to the intensity of the X-ray emission. The faintest objects have absorbed fluxes of $10^{-13}$ erg/cm$^2$/s while the brightest sources (No. 2 and 8) have about $5 \times 10^{-11}$ erg/cm$^2$/s.

Fig. 7 shows the broad band image (0.1–2.4 keV) of the GB 920622 location with the COMPTEL likelihood contours and the BATSE/Ulysses triangulation arc superimposed. In the following, we restrict ourselves to the best position from the combined COMPTEL/BATSE/Ulysses data.

For the optical identification we made use of the objective prism plate survey of Hamburg Observatory (Engels et al. 1988). These plates record 3500–5500 Å spectra with a dispersion of 1390 Å/mm down to 17–18th mag. In addition, direct plates with limiting magnitude 19–20 allow us to find fainter candidates when the prism plates are not conclusive, and these direct plates are also used to estimate the positions of the counterparts identified.

Table 5 shows all X-ray sources within an (arbitrary) range of $45'$ around the triangulation arc together with their optical identifications: column 1 gives a running number according to Fig. 7, column 2 the ROSAT source name, column 3 the ROSAT PSPC count rate in the 0.1–2.4 keV band, column 4 the optical identification, column 5 the optical brightness (typical error of $\pm 0.5$ mag for the faint candidates), and column 6 the optical position as measured on the corresponding direct plate with a typical error of $\pm 5$ arcsec. There is no X-ray source fully inside the Ulysses/BATSE/COMPTEL error box, in agreement with the above expectation. In the enlarged search area of $45'$ around the triangulation arc we find 10 X-ray sources while one would expect 14.

**Table 5.** ROSAT X-ray sources near the triangulation arc ($\pm 45'$)

| No. | Name | Count rate cts/sec | Identification [1] | Brightness $m_B$ (mag.) | Optical Position (2000.0) |
|---|---|---|---|---|---|
| 56 | RX J1036.6+4743 | 0.113 | HD 91782 | 8.7 | 10 36 48.1 +47 43 17 |
| 58 | RX J1040.2+4731 | 0.082 | Blue WK | 18.9 | 10 40 17.5 +47 30 59 |
| 60 | RX J1033.7+4718 | 0.026 | Blue WK | 18.6 | 10 33 46.2 +47 18 04 |
| 63 | RX J1045.0+4656 | 0.028 | EBL WK | 18.5 | 10 45 02.8 +46 56 55 |
| 66 | RX J1051.4+4644 | 0.022 | GSC star | 13.5 | 10 51 27.4 +46 44 39 |
| 67 | RX J1038.1+4642 | 0.024 | BL-WK | 18.5 | 10 38 08.8 +46 42 49 |
| 71 | RX J1052.5+4620 | 0.031 | GSC star | 12.9 | 10 52 30.5 +46 20 12 |
| 72 | RX J1044.0+4612 | 0.156 | HD 92855 | 7.9 | 10 44 01.8 +46 12 28 |
| 81 | RX J1050.6+4551 | 0.024 | Red WK[2] | 19.0 | 10 50 38.5 +45 51 19 |
| 82 | RX J1102.8+4542 | 0.044 | QSO[3], z=0.77 | 17.4 | 11 02 50.4 +45 42 31 |

[1] For explanation of abbreviations see text. [2] Identification uncertain. No other optical object down to 19th mag. [3] Redshift determined by optical spectroscopy (Bade et al. 1995).

All X-ray sources are rather faint with the total number of counts ranging between 10–60. Thus, no meaningful variability analysis could be performed. However, we have checked for any flaring emission of all sources of Table 5, but the distribution of photon arrival times is consistent with all sources being constant X-ray emitters.

Due to the high galactic latitude (BII=55–60°), most of these sources are identified with AGN candidates. Four objects are foreground (i.e. galactic) stars exhibiting coronal X-ray emission (including two Guide Star Catalog stars). Due to the low-dispersion spectra a definite identification is not always possible. In Table 5 the abbreviations "Blue WK" (weak source with blue continuum) denote possible AGN candidates and "EBL WK" (weak source with extreme blue continuum) very likely AGN candidates.

Having partly reliable counterparts and good candidates for the other X-ray sources we may state that there is no unusual X-ray source among those listed in Table 5. Since we know nothing about GRB counterparts, one is forced to search for unusual sources of any imaginable kind though it could also be a usual one. According to this criterion, we do not believe we have found a counterpart candidate for GB 920622. Thus, the steady soft X-ray flux of the source of GB 920622 is below the detection threshold for a point source, i.e. below 0.01 cts/sec (0.1–2.4 keV) in this case.

### 4.2. Simultaneous optical coverage

We are not aware of any simultaneous coverage of the GRB error box by regular sky patrols. For the European network (Greiner et al. 1994) the burst occurred after sunrise, and the Explosive Transient Camera on Kitt Peak was observing a different location (near the meridian) at the burst time (Vanderspek 1993).

### 5. Discussion

Observations by previous detectors (Norris et al. 1986) and the first results from the BATSE spectroscopy detectors (Band et al. 1992) have shown that the spectral shape of the $\gamma$-ray burst emission changes during typical bursts. In addition, spectral analysis of the COMPTEL burst module data of the bright burst 1B 910814B has shown that a break in the spectrum may move gradually to lower energies as the burst evolves (Hanlon et al. 1994). As a consequence, the shape of the spectrum of GB 920622 (Fig. 4) accumulated over the entire burst may well be affected by spectral evolution. At least in the COMPTEL data of GB 920622 there is evidence (Fig. 3) that spectral softening occurred within the first pulse of this burst. Similar hard to soft spectral evolution within individual burst pulses was reported by Norris et al. (1986) and very recently by Bhat et al. (1994) for a majority of bursts which were selected to have short rise times ($< 4$s) and a nearly exponential decay. Unfortunately, the separation of evolutionary effects is limited by the photon statistics at high (COMPTEL) and low (Ulysses) energies with the present detectors and not possible in the combined, broad-band analysis. In a recent time-resolved spectroscopy investigation of bright BATSE bursts, which included GB 920622, Ford et al. (1995) found a correlation between intensity

920622 this means that the maximum in $E_p$ is increasing in each of the successive pulses, i.e. in line with the maximum intensity of the pulses. It is mentioned again that there is no evidence at MeV energies (COMPTEL single instrument analysis) for spectral evolution from pulse to pulse.

The smoothly curved spectrum of GB 920622 as presented here as well as detailed analysis of BATSE SD data for a sample of bursts (Band et al. 1993) show that the introduction of a broken power law with a characteristic break energy for GRB spectra might be oversimplified.

Though the spectral model used (Band et al. 1993) does not imply a direct relation to a physical process, it describes the GRB spectrum over the entire energy range measured. It was noted (Band et al. 1993) that fitting the "GRB model" to the BATSE SD spectra generally leaves $\beta$ very uncertain. The inclusion of COMPTEL data removes this ambiguity (see Table 2) in the cases, where high signal-to-noise data are available up to $\approx 10$ MeV. Therefore, fits to the combined BATSE/COMPTEL data of selected GRBs are also well suited to constrain the shape and distribution of GRB high-energy tails (the $E^\beta$ component).

The absence of a well defined break can serve as additional evidence that the direct relation to a fundamental emission process (Baring 1992) may not be justified. Two-photon pair production (Schmidt 1978) is one of the most popular mechanisms invoked to explain the postulated spectral breaks (Baring 1992). However, a relativistic treatment of beaming (giving a relation between photon beaming angle, source distance and spectral break energy) reveals that the strong beaming required would blue-shift the $\gamma-\gamma$ attenuation breaks to energies much higher than 1 MeV (Baring 1994). This suggests that a mechanism other than pair production generates the spectral curvature even if one assumes that a certain parameter range for the emission conditions (e.g. density, magnetic and/or field, pressure etc.) is active during the burst or that these parameters change faster than our ability to measure significant spectra.

Current theories of the GRB emission process are not very constraining, partly because the models are controversial where the distance scale is concerned. Thus only few models are predictive enough to be testable using the spectrum of GB 920622. One of these is the blast wave model of Meszaros & Rees (1993). In short, if GRBs are at cosmological distances, the luminosity is so high that the resulting relativistic plasma expands with nearly the speed of light, producing a blast wave ahead of it and a reverse shock moving into the ejecta. Theoretical spectra have been calculated over a wide energy range (Meszaros & Rees 1993) as a function of magnetic field strength and particle acceleration mechanisms in the shocks. Though originally developed for cosmological bursts, a similar mechanism with scaled-down parameters may also be relevant the general form of these basic models (see Fig. 1 in Meszaros & Rees (1993)) is consistent at the low-energy end, these models predict a general flatter high-energy part ($> 1$ MeV) of the spectrum than observed in GB 920622. As another example out of the predictive models the original fireball scenario with its high temperature, modified blackbody spectrum (Goodman 1986) is clearly ruled out for GB 920622.

As far as the negative result of the search for quiescent X-ray sources is concerned, it is interesting to note that these ROSAT observations were performed 1.5 years before the $\gamma$-ray burst. It has been argued (Lasota 1992) that under the assumption of a slowly accreting neutron star as a $\gamma$-ray burst source, the burst could produce a shock which would prevent accretion onto the neutron star for a time span of several years following the burst. Thus, the above limits for X-ray emission from the GB 920622 source (0.01 cts/sec corresponds to $10^{-13}$ erg cm$^{-2}$ s$^{-1}$ for a $10^6$ K blackbody at $N_H = 10^{20}$ cm$^{-2}$ — which is the maximum galactic absorption at the GB 920622 position) constrains the pre-burst accretion rate of the neutron star to

$$\dot{M} \leq 1.5 \times 10^{-17} \left[\frac{R}{(10km)}\right] \left[\frac{M_{NS}}{M_\odot}\right]^{-1} \left[\frac{D}{(100pc)}\right]^2 M_\odot/yr.$$

Thus only for distances larger than $\approx 300$ pc would the accretion rate be high enough to trigger a hydrogen flash (Hameury et al. 1983).

## 6. Conclusions

The broad-band spectrum of GB 920622 as measured by Ulysses/BATSE/COMPTEL over the energy range 20 keV up to 10 MeV shows continuous curvature and is well fitted with the "GRB model" (Band et al. 1993). The composite spectrum has a sufficiently broad energy range (3 orders of magnitude) so that sophisticated physical models are necessary to explain the curvature.

There are no spectral features superimposed on the broad continuum of this burst. We find evidence for spectral evolution within the first pulse, but not from pulse to pulse of the burst. Interestingly, the mean hardness ratios of the three burst pulses are identical within the statistical errors.

In and around the $2° \times 1\rlap{.}'1$ error box of GB 920622 determined by COMPTEL and the BATSE/Ulysses triangulation arc we have investigated the quiescent X-ray sources found in the ROSAT All-Sky-Survey database. Optical follow-up observations of these X-ray sources have led to the identification of most of them. We believe that none of these X-ray sources is a quiescent X-ray counterpart candidate for GB 920622.

*Acknowledgements.* JG is extremely obliged to F. Haberl for extensive help with XSPEC (i.e. installing the Band model ting above 1 MeV). JG acknowledges help with the Ulysses data analysis by H. Richter and timely communication of the EGRET TASC data by G. Kanbach and E. Schneid. This research has made use of the Simbad database, operated at CDS, Strasbourg, France. JG is supported by the Deutsche Agentur für Raumfahrtangelegenheiten (DARA) GmbH under contract number FKZ 50 OR 9201. LH acknowledges The Education Programme of the European Community for a fellowship and ESA for providing facilities. The COMPTEL project is supported by DARA under contract No. 50 QV 90960, by NASA under contract NAS5−26645 and by the Netherlands Organisation for Scientific Research (NWO). The Ulysses GRB experiment was constructed at the Centre d'Etude Spatiale des Rayonnements in France with support from the Centre National d'Etudes Spatiales and at the Max-Planck Institute in Germany with support from FRG Contracts 01 ON 088 ZA/WRK 275/4-7.12 and 01 ON 88014. KH gratefully acknowledges assistance from JPL Contract 958056 and NASA Grant NAG5-1560.